\documentclass[onecol]{epl2}
\usepackage{subfigure}
\title{Distributed entanglement induced by dissipative bosonic media}
\shorttitle{Reference G28511} %Insert here a short version of the title if it exceeds 70 characters

\author{Li-Tuo Shen\inst{1} \and Xin-Yu Chen\inst{1} \and Zhen-Biao Yang\inst{2} \and Huai-Zhi
Wu\inst{1} \and Shi-Biao Zheng\inst{1\footnote{E-mail:
sbzheng11@163.com}}} \shortauthor{Li-Tuo Shen \etal}

\institute{
  \inst{1} Lab of Quantum Optics, Department of Physics, Fuzhou University, Fuzhou 350002, P. R. China\\
  \inst{2} Key Laboratory of Quantum Information, University
of Science and Technology of China, Chinese Academy of Sciences,
Hefei 230026, P. R. China }


\abstract{We describe a scheme with analytic result that allows to
generate steady-state entanglement for two atoms over a dissipative
bosonic medium. The resonant coupling between the mediating bosonic
mode and cavity modes produces three collective atomic decay
channels. This dissipative dynamics, together with the unitary
process induced by classical microwave fields, drives the two atoms
to the symmetric or asymmetric entangled steady state conditional
upon the choice of the phases of the microwave fields. The effects
on the steady-state entanglement of off-resonance mediating bosonic
modes are analyzed. The entanglement can be obtained with high
fidelity regardless of the initial state and there is a linear
relation in the scaling of the fidelity with the cooperativity
parameter. The fidelity is insensitive to the fluctuation of the
Rabi frequencies of the classical driving fields.}

\begin{document}
\maketitle

\section{Introduction}
Quantum entanglement plays a crucial role in performing quantum
information processing
\cite{PRL2003-91-097905,PRL2000-85-2392,PRL2000-85-1762,APB-1995-60-129},
and it is the basic requirement for quantum communication to
transfer quantum state between distant locations. To construct
quantum networks, one needs to generate spatially separate qubits,
store them for sufficiently long time, and perform logic operations
between them. Therefore, the generation of entanglement between
different nodes becomes an important task \cite{PRL-2003-91-110405}.
The optical absorption and other channel noise inevitably bring the
detrimental effects to the entanglement between distant nodes, which
normally decreases exponentially with the length of the connecting
medium. The quantum communication difficulty cannot be solved just
based on unitary dynamics
\cite{PRL2006-96-010503,JPB2010-43-085506,PRA2007-75-012324,EPJD2008-50-91,
PRA2010-82-042327,PRA2010-82-012307,PRA2008-78-063805,LPR2008-2-527,epjd2011-61-737}.

To overcome the problem associated with the exponential fidelity
decay of distributed entanglement, the concept of dissipative
dynamics can be used, which utilizes the dissipation as a resource
for quantum state engineering
\cite{PRL2011-106-090502,arXiv:1110.1024v1,arXiv:1112.2806v1,PRA2011-84-064302,PRA2011-84-022316,
PRL2011-106-020504,PRA2011-83-042329,PRL2011-107-120502,
PRA2010-82-054103,arXiv1005.2114v2,PRL2003-91-070402,arXiv:1202.0345v1,PRA2012-85-042320,JOSAB2011-28-228,PRA2012-85-022324,
EPL-85-20007,PRL-89-277901,PRA-76-022312,PRE-77-011112,PRA-77-042305,PRL-100-220401,arXiv:1110.6718v1,PRA-80-2009-022119}.
Recently, experimental realization of dissipative state preparation
has also been reported \cite{PRL2011-107-080503}. Here, we describe
a scheme with analytic result that allows to generate steady-state
entanglement for two atoms over a dissipative bosonic medium.
Compared with the previous scheme we proposed in Ref.
\cite{PRA2011-84-064302}, the steady entanglement can be produced
over a longer distance via a dissipative mediating mode, and is very
useful for testing quantum nonlocality \cite{PRL-1991-66-252},
quantum secret sharing \cite{PRA-1999-59-1829} and quantum
communication \cite{PRL-1997-78-3221}.

Due to the coherent cavity-medium coupling, the system becomes
mathematically equivalent to the case that two distant atoms
collectively interact with three nondegenerate delocalized field
modes, one mode inducing the asymmetric collective atomic decay
channel and the other two inducing the symmetric collective atomic
decay channels. The decay channels induced by three delocalized
field modes are analytically resolved with the dissipative dynamics.
Analytical result indicates that the distributed entanglement can be
obtained with high fidelity requiring neither the unitary feedback
control nor the photon detection. The steady-state entanglement
arises from the competition between the collective atomic decays and
the unitary evolution induced by the classical microwave fields,
which act as a push button to start the dynamical process. We find
that the scaling of fidelity $F$ with the cooperativity parameter
$C$ is linear. Besides, the effect on the steady-state entanglement
of dispersive mediating modes analyzed. We find that the couplings
between the dispersive mediating modes and cavity modes lead to the
frequency shifts of the normal delocalized bosonic modes, when one
of these modes is resonant with the classical laser field, the Raman
transitions between two atomic ground states dominate the atom-field
coupling dynamics and destroy the process of the steady-state
entanglement preparation. The steady-state entanglement is
faithfully prepared when the frequency space of the mediating
bosonic modes is large enough, showing that our scheme can work well
even when many mediating bosonic modes are involved.

\section{The theoretical model}
As shown in Fig. 1, two three-level atoms are individually trapped
in two single-mode cavities linked by a dissipative bosonic medium
(a third cavity or a waveguide). Each coherent driving $\Omega$
pumps the ground state $|0\rangle$ to the excited state $|2\rangle$
with the detuning $\Delta$. Atom-cavity interaction $g$ drives the
transition $|1\rangle$ $\leftrightarrow$ $|2\rangle$ with the
detuning $\Delta-\delta$, and two ground states are coupled by a
resonant microwave $\Omega_{M}$ in each cavity.

We assume two cavity modes are coupled to the mediating mode. Then
the Hamiltonian of the whole system in the interaction picture can
be written as $H_{I}$ = $H_{0}$ + $H_{g}$ + $V_{+}$ + $V_{-}$:

\begin{figure}
\centering
\includegraphics[width=1\columnwidth]{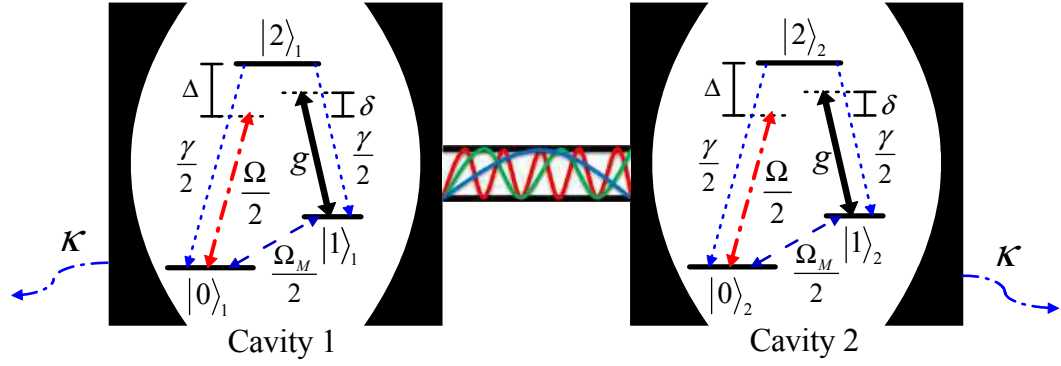} \caption{(Color
online) Experimental setup and level diagram for dissipative
preparation of long-distance entanglement between two atoms coupled
to two cavities respectively, which are connected by a multi-mode
dissipative medium.}
\end{figure}

\begin{eqnarray}\label{e1-e4}
H_{0}&=&\sum_{i=1}^{2} (\Delta|2\rangle_{i}\langle2|+\delta
a_{i}^{\dag}a_{i})+\sum_{i=1}^{2}
(g|2\rangle_{i}\langle1|a_{i}+H.c.)\cr&&
+\sum_{n}^{N}\bigg\{(\Delta_{n}+\delta)b_{n}^{\dagger}b_{n}+[\nu_{n}
b_{n}(a_{1}^{\dag}+a_{2}^{\dag})+H.c.]\bigg\},\cr&&\\
H_{g}&=&\frac{\Omega_{M}}{2}(|1\rangle_{1}\langle0|+e^{-i\theta_{M}}|1\rangle_{2}\langle0|)+H.c.,\\
V_{+}&=&\frac{\Omega}{2}(|2\rangle_{1}\langle0|+|2\rangle_{2}\langle0|),\\
V_{-}&=&\frac{\Omega}{2}(|0\rangle_{1}\langle2|+|0\rangle_{2}\langle2|),
\end{eqnarray}
where $a_{i}$ ($i=1,2$) is the annihilation operator for the $i$th
cavity field mode, $b_{n}$ is the annihilation operator for the
$n$th mediating mode, $\Delta_{n}$ denotes the frequency difference
between the $n$th mediating mode and the cavity mode. $\nu_{n}$
denotes the coupling strength between the $i$th cavity mode and the
$n$th mediating mode, $g$ is the atom-cavity coupling constant,
$\Omega$ and $\Omega_{M}$ represent the classical laser driving
strength and the microwave driving strength, respectively.
$\theta_{M}$ is the phase difference between the two microwave
fields applied to atoms $1$ and $2$. The cavity-mediating couplings
are assumed to be the same for all the mediating modes, i.e.,
$\nu_{n}$ $=$ $\nu$.

To analyze the main dissipation mechanism easily, we first focus on
the case that only one mediating mode (i.e., the first mediating
mode $b_{1}$) resonantly interacts with the cavity mode and the
other mediating modes with large detuning can be neglected in
$H_{0}$. Let us define the symmetric steady-state $|T\rangle$ $=$
$\big(|10\rangle+|01\rangle\big)/\sqrt{2}$, the asymmetric
steady-state $|S\rangle$ $=$
$\big(|10\rangle-|01\rangle\big)/\sqrt{2}$, and introduce three
normal delocalized bosonic modes as $c_{1}$ $=$
$(a_{1}-a_{2})/\sqrt{2}$, $c_{2}$ $=$
$(a_{1}+a_{2}+\sqrt{2}b_{1})/2$, and $c_{3}$ $=$
$(a_{1}+a_{2}-\sqrt{2}b_{1})/2$. In terms of the new operators, the
Hamiltonian $H_{0}$ becomes
\begin{eqnarray}\label{e5}
H_{0}&=&
\Delta(|2\rangle_{1}\langle2|+|2\rangle_{2}\langle2|)\cr\cr&&
+\bigg[g|2\rangle_{1}\langle1|(\frac{1}{2}c_{2}+\frac{1}{2}c_{3}+\frac{\sqrt{2}}{2}c_{1})\cr\cr&&
+g|2\rangle_{2}\langle1|(\frac{1}{2}c_{2}+\frac{1}{2}c_{3}-\frac{\sqrt{2}}{2}c_{1})+H.c.\bigg]\cr\cr&&
+\delta
c_{1}^{\dagger}c_{1}+(\delta+\sqrt{2}\nu)c_{2}^{\dagger}c_{2}+(\delta-\sqrt{2}\nu)c_{3}^{\dagger}c_{3}.
\end{eqnarray}
Here, $H_{0}$ describes the asymmetric coupling for two atoms to the
delocalized field mode $c_{1}$, and the symmetric couplings to
$c_{2}$ and $c_{3}$. Because of the cavity-medium coupling, all the
delocalized field modes are nondegenerate and each causes a
collective dissipation channel. $\kappa_{i}$ $(i=1,2)$ and
$\kappa_{3}$ represent the photon decay rate of cavity $i$ and the
medium, respectively. $\gamma_{0}$ and $\gamma_{1}$ denote the
atomic spontaneous emission rates for $|2\rangle$ $\rightarrow$
$|0\rangle$ and $|2\rangle$ $\rightarrow$ $|1\rangle$, respectively.
For simplicity we here set $\kappa_{1}$ $=$ $\kappa_{2}$ $=$
$\kappa_{3}$ $=$ $\kappa$ and $\gamma_0$ = $\gamma_1$ $=$
$\gamma/2$. Then the corresponding Lindblad operators associated
with the photon decay and atomic spontaneous emission can be
expressed as $L^{\kappa_1}$ = $\sqrt{\kappa}$ $c_1$, $L^{\kappa_2}$
= $\sqrt{\kappa}$ $c_2$, $L^{\kappa_3}$ = $\sqrt{\kappa}$ $c_3$,
$L^{\gamma_1}$ = $\sqrt{\gamma_0}$ $|0\rangle_{1}\langle2|$,
$L^{\gamma_2}$ = $\sqrt{\gamma_0}$ $|0\rangle_{2}\langle2|$,
$L^{\gamma_3}$ = $\sqrt{\gamma_1}$ $|1\rangle_{1}\langle2|$,
$L^{\gamma_4}$ = $\sqrt{\gamma_1}$ $|1\rangle_{2}\langle2|$.

Under the condition that the classical laser fields are weak, we can
apply the method based on the second-order perturbation theory to
tailor the effective dissipation process
\cite{PRL2011-106-090502,arXiv:1110.1024v1,arXiv:1112.2806v1}. Then
the dynamics of our distributed cavity QED system is governed by the
effective Hamiltonian $H_{eff}$ and effective Lindblad operator
$L_{eff}^{x}$:
\begin{eqnarray}\label{e6-e7}
H_{eff}&=&-\frac{1}{2}V_{-}[H^{-1}_{NH}+(H^{-1}_{NH})^{\dag}]V_{+}+H_{g},\\
L_{eff}^{x}&=&L^{x}H^{-1}_{NH}V_{+},
\end{eqnarray}
where $H^{-1}_{NH}$ is the inverse of the non-Hermitian Hamiltonian
$H_{NH}$ = $H_{0}$ $-$ $\frac{i}{2}\sum_{x}(L^{x})^{\dag}L^{x}$, and
its elements determine the strength of effective evolution process.
The resulting effective master equation in Lindblad form becomes
\begin{eqnarray}\label{e8-e12}
\dot{\rho}&=&i[\rho,H_{eff}]+\sum_{x}\{L_{eff}^{x}\rho
(L_{eff}^{x})^{\dag}\cr&&-\frac{1}{2}[(L_{eff}^{x})^{\dag}L_{eff}^{x}\rho+\rho(L_{eff}^{x})^{\dag}L_{eff}^{x}]\},\\
H_{eff}&=&-Re[\Omega^2\widetilde{R}_{1}]|00\rangle\langle
00|-Re[\frac{\Omega^2}{4}\widetilde{R}_{2}]|S\rangle\langle
S|\cr&&-Re[\frac{\Omega^2}{4}\widetilde{R}_{3}]|T\rangle\langle T|+H_{g},\\
L_{e}^{\kappa_{1}}&=&\sqrt{\frac{(\delta^2-2\nu^2)^2g_{e}^2\kappa/4}{A^2+B^2}}|S\rangle\langle00|\cr&&
+\sqrt{\frac{g_{e}^2\kappa/8}{C_{1}^2+D_{1}^2}}|11\rangle\langle
S|,\\
L_{e}^{\kappa_{2}}&=&\sqrt{\frac{\delta^2(\delta-\sqrt{2}\nu)^2g_{e}^2\kappa/8
}{A^2+B^2}}|T\rangle\langle00|
\cr&&+\sqrt{\frac{(\delta-\sqrt{2}\nu)^2g_{e}^2\kappa/16}{C_{2}^2+D_{2}^2}}|11\rangle\langle
T|,\\
L_{e}^{\kappa_{3}}&=&\sqrt{\frac{\delta^2(\delta+\sqrt{2}\nu)^2g_{e}^2\kappa/8
}{A^2+B^2}}|T\rangle\langle00|
\cr&&+\sqrt{\frac{(\delta+\sqrt{2}\nu)^2g_{e}^2\kappa/16}{C_{2}^2+D_{2}^2}}|11\rangle\langle
T|,
\end{eqnarray}
where $Re[$ $]$ denotes the real part of the argument,
\begin{eqnarray}\label{e13}
g_{e}&=&g\Omega,
\delta^{'}=\delta-i\kappa/2,\Delta^{'}=\Delta-i\gamma/2,\cr
\widetilde{R}_{1}&=&\frac{\delta^{'}(\delta^{'2}-2\nu^2)}{\Delta^{'}\delta^{'}(\delta^{'2}-2\nu^2)-g^2(\delta^{'2}-\nu^2)},\cr
\widetilde{R}_{2}&=&\frac{\Delta^{'}\delta^{'}(\delta^{'2}-2\nu^2)-g^2\delta^{'2}}{(g^2-\delta^{'}\Delta^{'})(\delta^{'2}\Delta^{'}-\delta^{'}g^2+2\Delta^{'}\nu^2)},\cr
\widetilde{R}_{3}&=&\frac{\Delta^{'}\delta^{'}(\delta^{'2}-2\nu^2)-g^2(\delta^{'2}-2\nu^2)}{(g^2-\delta^{'}\Delta^{'})(\delta^{'2}\Delta^{'}-\delta^{'}g^2+2\Delta^{'}\nu^2)},\cr
A&=&\Delta\delta(\delta^2-2\nu^2)-g^2(\delta^2-\nu^2),\cr
 B&=&(\delta/2-\nu^2)(\Delta\kappa+\gamma\delta)+\delta\kappa(\Delta\delta-g^2),\cr
 C_{1}&=&g^2-\Delta\delta, \cr
 D_{1}&=&(\Delta\kappa+\delta\gamma)/2,\cr
 C_{2}&=&g^2\delta-\Delta(\delta^2-2\nu^2),\cr
 D_{2}&=&\kappa(\Delta\delta-g^2/2)+\gamma(\delta^2-2\nu^2)/2.
\end{eqnarray}
Since $\Omega$ is very small, we can neglect those terms containing
the effective shifts $\mathcal {O}(\Omega^2)$ in Eq. (9), so that
$H_{eff}$ $\simeq$ $H_g$. There are three primarily effective decay
channels characterized by $L_{e}^{\kappa_1}$, $L_{e}^{\kappa_2}$,
$L_{e}^{\kappa_3}$ through three delocalized bosonic modes $c_{1}$,
$c_{2}$, $c_{3}$, respectively. It is the mediating mode that links
the two separate field modes and lifts the degeneracy of three
delocalized field modes, leading to three independent decay
channels. We assume $\kappa$, $\gamma$ $\ll$ $g$, $\delta$, $\nu$,
$\Delta$, so that the minor terms with higher order than $\kappa^2$,
$\gamma^2$ and $\kappa\gamma$ in the denominators of
$L_{e}^{\kappa_{1}}$, $L_{e}^{\kappa_{2}}$ and $L_{e}^{\kappa_{3}}$
have been omitted.

\begin{figure}
\centering
\includegraphics[width=1\columnwidth]{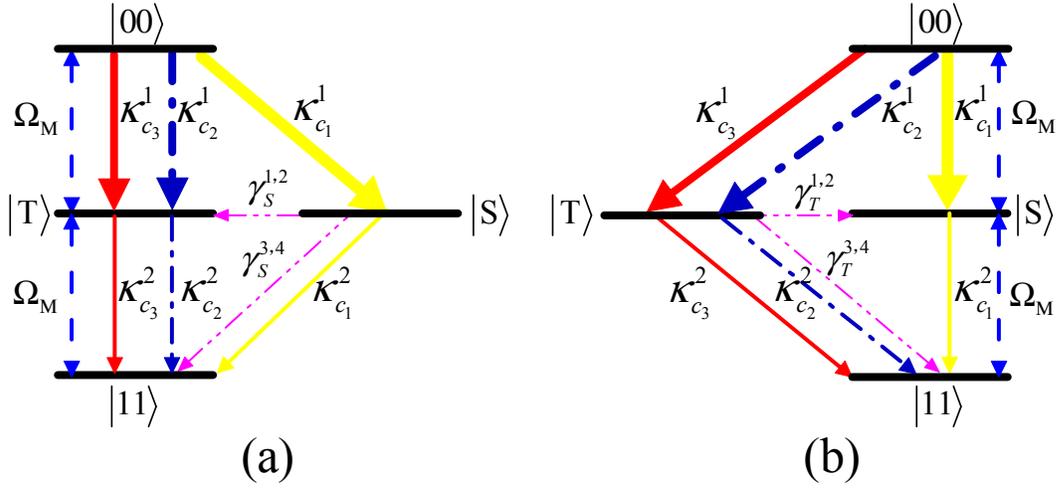} \caption{(Color
online) Effective two qubit system for coherent and dissipative
processes in the ground state basis $\big\{$ $|00\rangle$,
$|S\rangle$, $|T\rangle$, $|11\rangle$ $\big\}$. Effective
dissipative rates induced by the field decay from $|00\rangle$ to
$|S\rangle$ and $|T\rangle$ are $\kappa_{c_{1}}^1$ and
($\kappa_{c_{2}}^1$ $+$ $\kappa_{c_{3}}^1$), and from $|S\rangle$
and $|T\rangle$ to $|11\rangle$ are $\kappa_{c_{1}}^2$ and
($\kappa_{c_{2}}^2$ $+$ $\kappa_{c_{3}}^2$). $\gamma_{S}^{1,2}$,
$\gamma_{S}^{3,4}$, $\gamma_{T}^{1,2}$, and $\gamma_{T}^{3,4}$ are
the effective dissipative rates due to atomic spontaneous emission.
(a) $\theta$ $=$ $\pi$; (b) $\theta$ $=$ $0$.}
\end{figure}

As shown in Figs. 2(a) and 2(b), $L_{e}^{\kappa_{1}}$ indicates the
effective decay from $|00\rangle$ to $|S\rangle$ at a rate
$\kappa_{c_{1}}^{1}$ and from $|S\rangle$ to $|11\rangle$ at a rate
$\kappa_{c_{1}}^{2}$, which is caused by $c_{1}$ mode that only
contains the contribution of cavity mode. $L_{e}^{\kappa_{2}}$
($L_{e}^{\kappa_{3}}$) indicates the effective decay from
$|00\rangle$ to $|T\rangle$ at a rate $\kappa_{c_{2}}^{1}$
($\kappa_{c_{3}}^{1}$) and from $|T\rangle$ to $|11\rangle$ at a
rate $\kappa_{c_{1}}^{2}$ ($\kappa_{c_{3}}^{2}$), which is caused by
$c_{2}$ ($c_{3}$) mode that contains the contributions of both the
mediating mode and cavity modes. Three effective collective decays
happen simultaneously. The decay rates $\kappa_{c_{1}}^{1}$
($\kappa_{c_{1}}^{2}$), $\kappa_{c_{2}}^{1}$ ($\kappa_{c_{2}}^{2}$)
and $\kappa_{c_{3}}^{1}$ ($\kappa_{c_{3}}^{2}$) equal to the square
of the first (second) coefficient in the right hand side of Eq.
(10), Eq. (11), and Eq. (12), respectively. To prepare the desired
steady-state entanglement $|S\rangle$ ($|T\rangle$), it is necessary
to suppress the effective decays from $|S\rangle$ to $|11\rangle$
(from $|T\rangle$ to $|11\rangle$), i.e., the condition
$\kappa_{c_{1}}^{1}$ $\gg$ $\kappa_{c_{1}}^{2}$
($\kappa_{c_{2}}^{1}$ $\gg$ $\kappa_{c_{2}}^{2}$ and
$\kappa_{c_{3}}^{1}$ $\gg$ $\kappa_{c_{3}}^{2}$) should be
satisfied. Two microwave fields cause the coherent shuffling of the
three states $|00\rangle$, $|T\rangle$ ($|S\rangle$) and
$|11\rangle$ for $\theta_{M}$ $=$ $0$ ($\pi$). As a result, the
state $|S\rangle$ ($|T\rangle$) is the unique stationary point of
the system.

\section{The fidelity}
Using Eq. (7), we obtain the effective Lindblad operators that drive
the population out of the target state $|S\rangle$
\begin{eqnarray}\label{e14-e15}
L_{e,S}^{\gamma_{1}}&=&L_{e,S}^{\gamma_{2}}=\sqrt{\gamma_{S}^{1,2}}|T\rangle\langle
S|,\\
L_{e,S}^{\gamma_{3}}&=&L_{e,S}^{\gamma_{4}}=\sqrt{\gamma_{S}^{3,4}}|11\rangle\langle
S|,
\end{eqnarray}
and those driving the population out of the target state $|T\rangle$
\begin{eqnarray}\label{e16-e17}
L_{e,T}^{\gamma_{1}}&=&L_{e,T}^{\gamma_{2}}=\sqrt{\gamma_{T}^{1,2}}|S\rangle\langle
T|,\\
L_{e,T}^{\gamma_{3}}&=&L_{e,T}^{\gamma_{4}}=\sqrt{\gamma_{T}^{3,4}}|11\rangle\langle
T|,
\end{eqnarray}
where $\gamma_{e}$ = $\gamma\Omega^2\nu^4$ $/$ $\big[$
$\Delta(\delta^2-\nu^2)+\delta g^2$$\big]^2$, $\gamma_{S}^{1,2}$ $=$
$\gamma_{T}^{1,2}$ $=$ $\gamma_{e}/32$ and $\gamma_{S}^{3,4}$ $=$
$\gamma_{T}^{3,4}$ $=$ $\gamma_{e}/16$. The evolutions of
populations of different state components governed by the full
master equation are plotted in Fig. 3, which shows that the
steady-state entanglement $|S\rangle$ or $|T\rangle$ can be obtained
with high fidelity depending upon the choice of the phase difference
$\theta_{M}$ between the two microwave fields.

\begin{figure}\label{fig-3}
\centering \subfigure[]{ \label{Fig.sub.a}
\includegraphics[width=0.95\columnwidth]{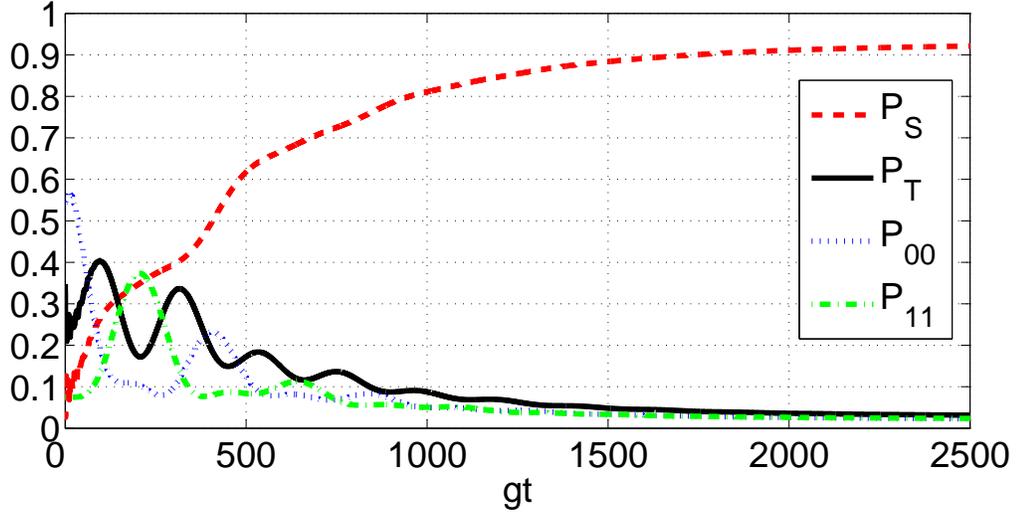}}\\
\subfigure[]{ \label{Fig.sub.b}
\includegraphics[width=0.95\columnwidth]{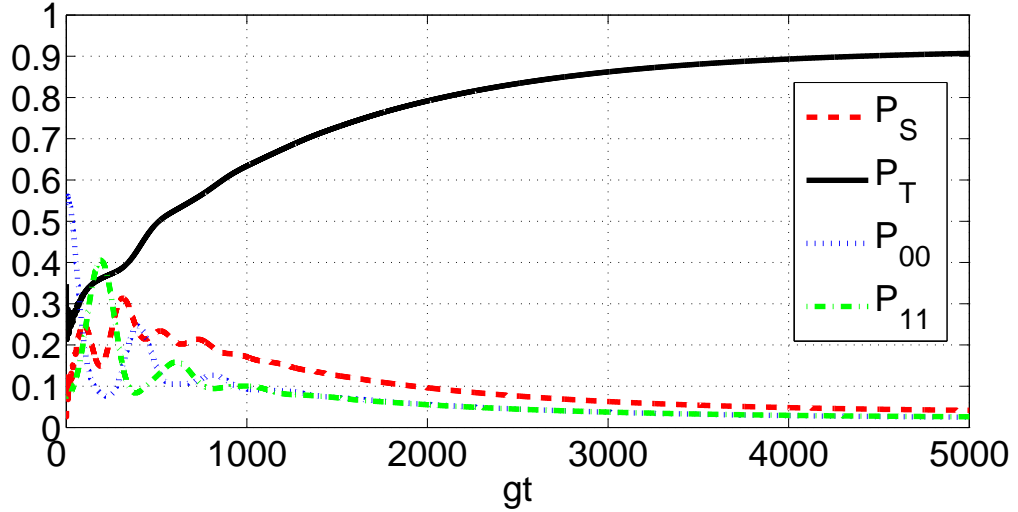}}
\caption{(Color online) The populations of states $|S\rangle$,
$|T\rangle$, $|00\rangle$, and $|11\rangle$ versus the dimensionless
parameter $gt$ for a random initial state by solving the full master
equation. The curves are both plotted for an appropriate set of
parameters $C$ $=$ $150$, $\gamma$ $=$ $2\kappa$, $\kappa=0.0577g$,
$\gamma=0.1154g$, $\Omega$ = $0.06g$, $\Omega_{M}$ $=$ $0.0138g$,
$\Delta$ $=$ $1.3g$, $\nu$ $=$ $0.4528g$ and $\delta$ $=$ $0.2875g$.
(a) $\theta_{M}$ $=$ $0$; (b) $\theta_{M}$ $=$ $\pi$.}
\end{figure}

\begin{figure}\label{fig-4}
\centering
\includegraphics[width=0.9\columnwidth]{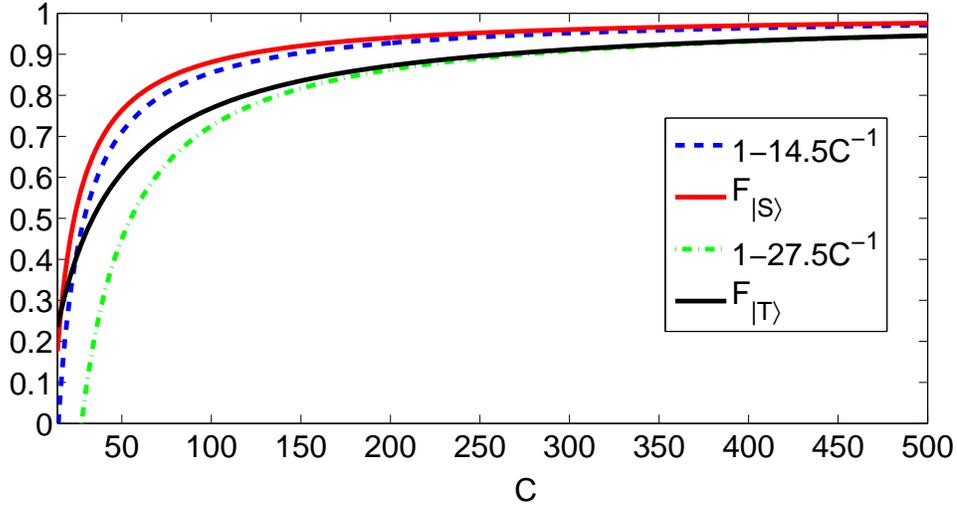} \caption{(Color
online) The fidelity for steady-state entanglements $|S\rangle$ (red
solid line) and $|T\rangle$ (black solid line) vs the cooperativity
parameter $C$, and the mathematical fitting curves for maximizing
$F_{|S\rangle}$ (blue dot line) and $F_{|T\rangle}$ (green dot dash
line).}
\end{figure}

To evaluate the fidelity of the steady-state $|S\rangle$ or
$|T\rangle$, we apply the rate equation as follows
\begin{eqnarray}\label{e18-e19}
\dot{P}_{S}&=&\kappa_{c_{1}}^{1}P_{00}-
\big(\kappa_{c_{1}}^{2}+\gamma_{S}^{1,2}+\gamma_{S}^{3,4}\big)P_{S},\\
\dot{P}_{T}&=&\big(\kappa_{c_{2}}^{1}+\kappa_{c_{3}}^{1}\big)P_{00}\cr&&
-\big(\kappa_{c_{2}}^{2}+\kappa_{c_{3}}^{2}+\gamma_{T}^{1,2}+\gamma_{T}^{3,4}\big)P_{T},
\end{eqnarray}
where $P_{\varphi}$ is the probability to be in the state
$|\varphi\rangle$. When the whole system reaches the steady
entanglement $|\varphi\rangle$, we assume $P_{\varphi}$ $\simeq$ $1$
and the probability in each of the other three states is nearly
$P_{00}$, then $1-F_{|B\rangle}$ $\approx$ $3P_{00}$ ($B$ $=$ $S$,
$T$). Here, $F_{|B\rangle}$ $=$ $|\langle B|\rho|B\rangle|$ is the
fidelity of state $B$, and
\begin{eqnarray}\label{e20-e21}
1-F_{|S\rangle}&\approx&[\frac{3g_{e}^2\kappa}{8(C_{1}^2+D_{1}^2)}+\frac{9\gamma_{e}}{32}]
\big/[\frac{(\delta^2-2\nu^2)^2g_{e}^2\kappa}{4B^2}],\cr\cr&&
\\
1-F_{|T\rangle}&\approx&[\frac{3(\delta^2+2\nu^2)g_{e}^2\kappa}{8(C_{2}^2+D_{2}^2)}+\frac{9\gamma_{e}}{32}]
\big/[\frac{\delta^2(\delta^2+2\nu^2)g_{e}^2\kappa}{4B^2}].\cr\cr&&
\end{eqnarray}
We plot the fidelity for entangled steady-states with the
cooperativity parameter $C$ and gain a simplified formula through
mathematical fitting based on the least squares algorithm in Fig. 4,
and find out the actual constants for maximizing the fidelity as
follows
\begin{eqnarray}\label{e22-e23}
1-F_{|S\rangle}&\simeq&14.5C^{-1},\\
 1-F_{|T\rangle}&\simeq&27.5C^{-1}.
\end{eqnarray}

The fidelities of the steady-state versus the fluctuations of
coherent driving $\Omega$ and microwave $\Omega_{M}$ are plotted
with full master equation in Figs. 5(a) and 5(b), and the results
shows that the fidelity $F_{|S\rangle}$ and $F_{|T\rangle}$ remain
higher than $91\%$ even when the relative errors in $\Omega$ and
$\Omega_{M}$ reach $20\%$.

\begin{figure}\label{fig-5}
\centering \subfigure[]{ \label{Fig.sub.a}
\includegraphics[width=0.45\columnwidth]{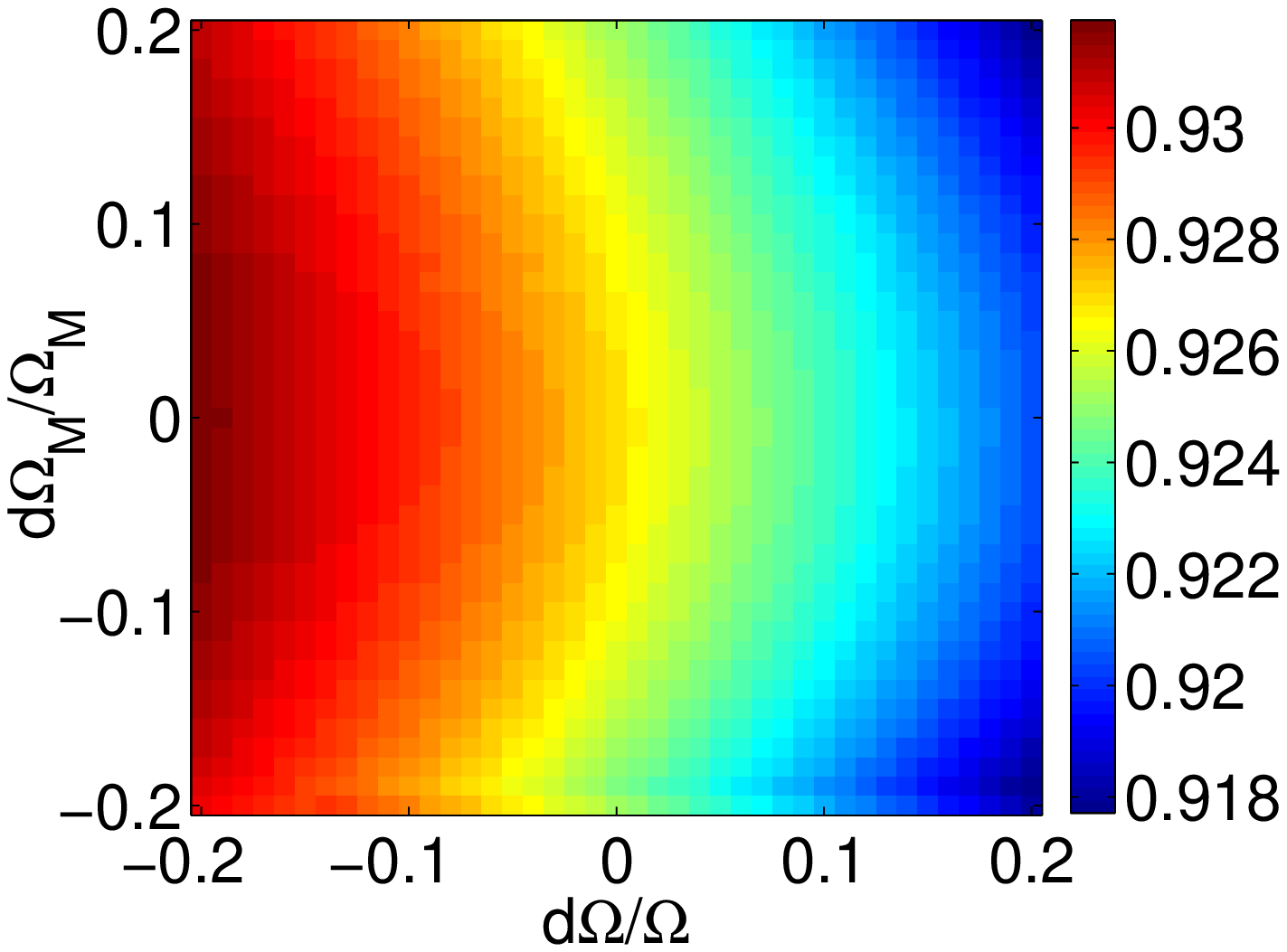}}
\subfigure[]{ \label{Fig.sub.b}
\includegraphics[width=0.45\columnwidth]{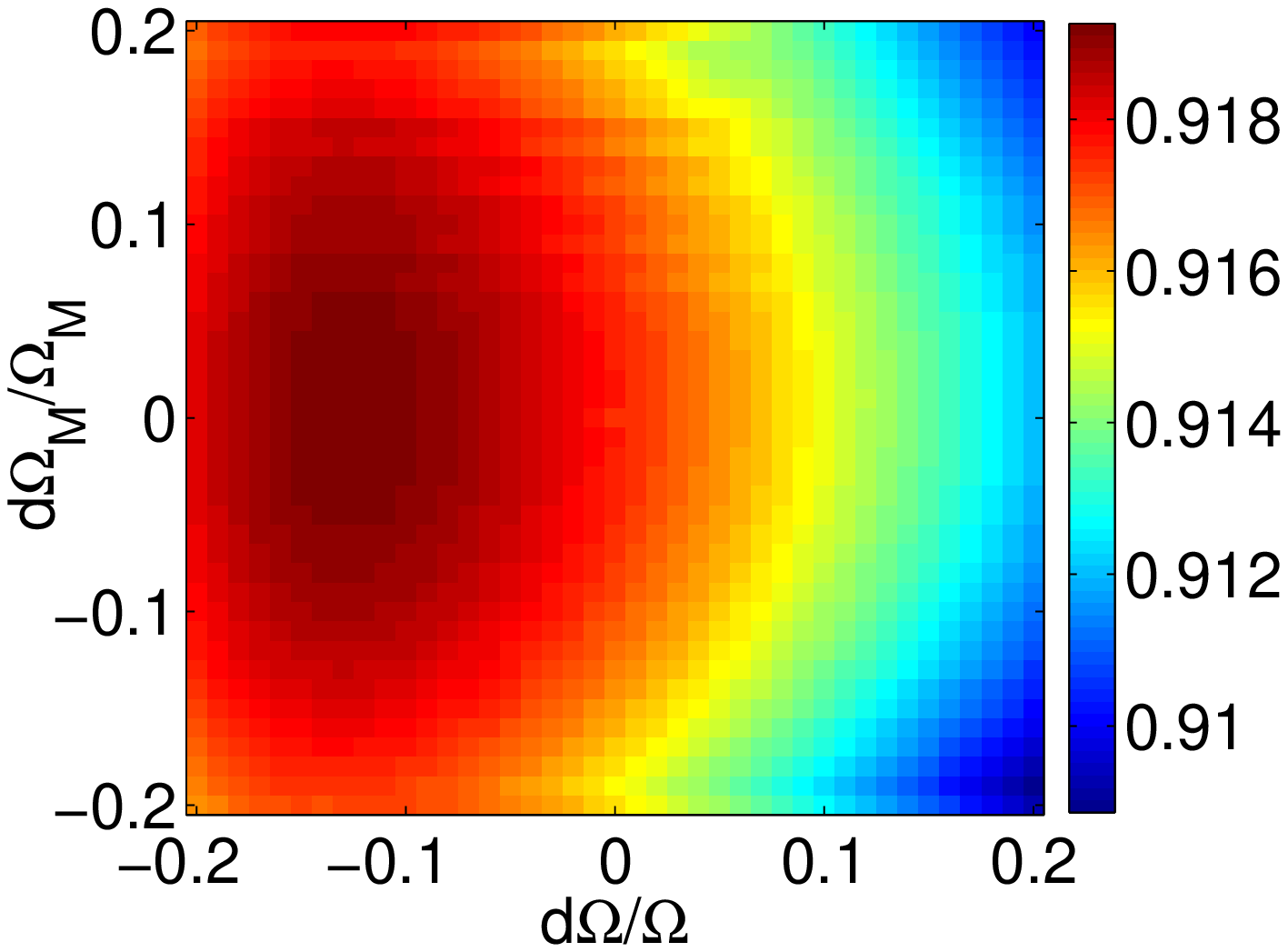}}
\caption{(Color online) Fidelity of the steady states versus the
fluctuations of $\Omega$ and $\Omega_{M}$. $d\Omega$ and
$d\Omega_{M}$ are the corresponding deviations of $\Omega$ and
$\Omega_{M}$, respectively. Both plots are calculated with full
master equation at time $t$ $=$ $9000$. (a) $F_{|S\rangle}$ vs
$d\Omega/\Omega$ and $d\Omega_{M}/\Omega_{M}$; (b) $F_{|T\rangle}$
vs $d\Omega/\Omega$ and $d\Omega_{M}/\Omega_{M}$.}
\end{figure}

\section{Influences of the dispersive mediating bosonic modes}
In the above derivations, we assume that only one resonant mediating
mode interacts with the cavity modes. We now consider the effects on
the steady-state entanglement of other existing dispersive bosonic
modes. Fig. 6 shows influences of the number $N$ of the mediating
modes and the frequency spacing $\Delta_{x}$ on the fidelity
$|F\rangle_{S}$.

$F_{|T\rangle}$ is influenced severely by the dispersive mediating
modes and it is difficult to obtain the steady-state $|T\rangle$
when $N$ becomes large. Therefore, we just discuss the state
$|S\rangle$ as follows. We first consider the case that two
mediating modes are involved in Fig. 6(a), where we assume the first
mediating mode is resonant with the cavity modes. There is a dip
when $\Delta_{x}$ $=$ $-0.64g$. This can be explained as follows.
The coupling between the mediating modes and cavity modes leads to
frequency shifts of the delocalized field modes. When one of the
delocalized modes, together with the classical laser field happens
to induce the resonant Raman transition between two ground states
($|0\rangle$ and $|1\rangle$), as shown in Fig. 7, the Raman
process, together with cavity decay, leads to effective decay
$|0\rangle$ $\rightarrow$ $|1\rangle$ and destroys the entanglement.

The fidelity versus frequency spacing with three (five) mediating
modes is plotted in Fig. 6(b) ( Fig. 6(c) ), showing that two
valleys appear when $\Delta_{x}$ $=$ $-0.54g$ ($-0.58g$) and
$\Delta_{x}$ $=$ $0.54g$ ($0.58g$). The appearance of two symmetric
valleys in Fig. 6(b) ( Fig. 6(c) ) is due to the fact that two
(four) equally spaced mediating modes are symmetrically distributed
in two sides of the resonant mode. Especially, two small valleys
appear when $|\Delta_{x}|$ $=$ $0.20g$ in Fig. 6(c), where two of
those delocalized field modes are degenerate.

\begin{figure}\label{fig-6}
\centering \subfigure[]{ \label{Fig.sub.a}
\includegraphics[width=0.75\columnwidth]{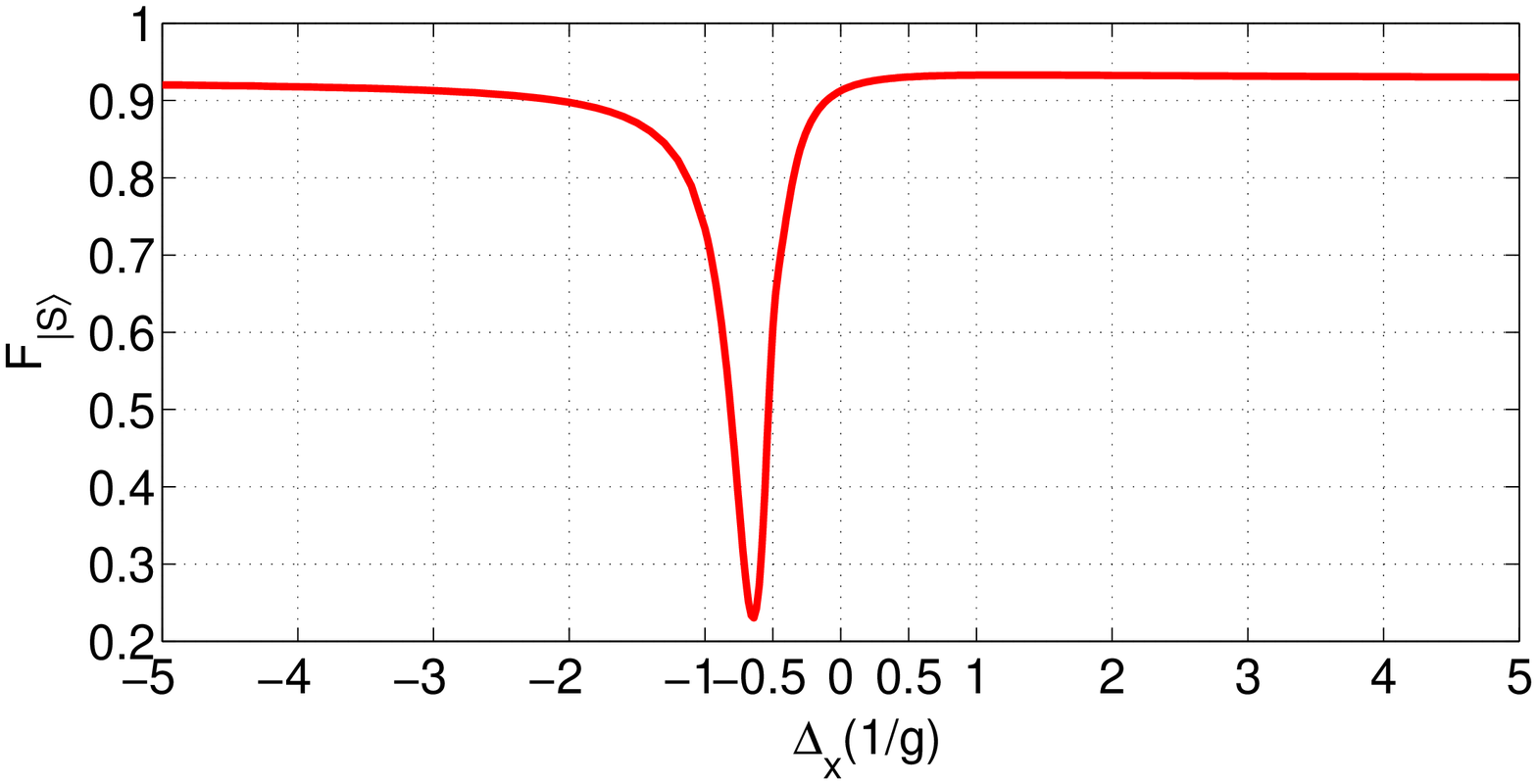}}
\subfigure[]{ \label{Fig.sub.b}
\includegraphics[width=0.75\columnwidth]{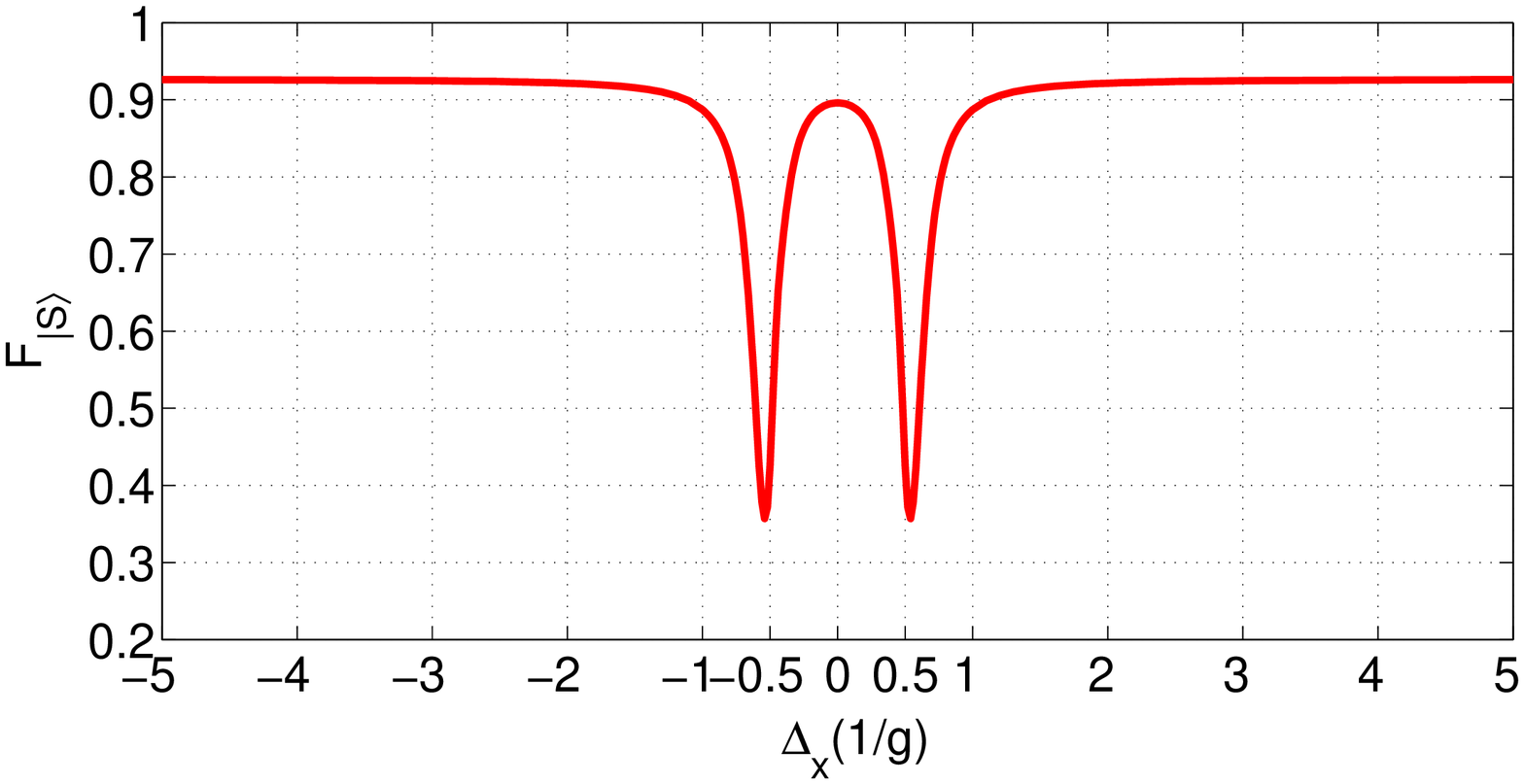}}
\subfigure[]{ \label{Fig.sub.c}
\includegraphics[width=0.75\columnwidth]{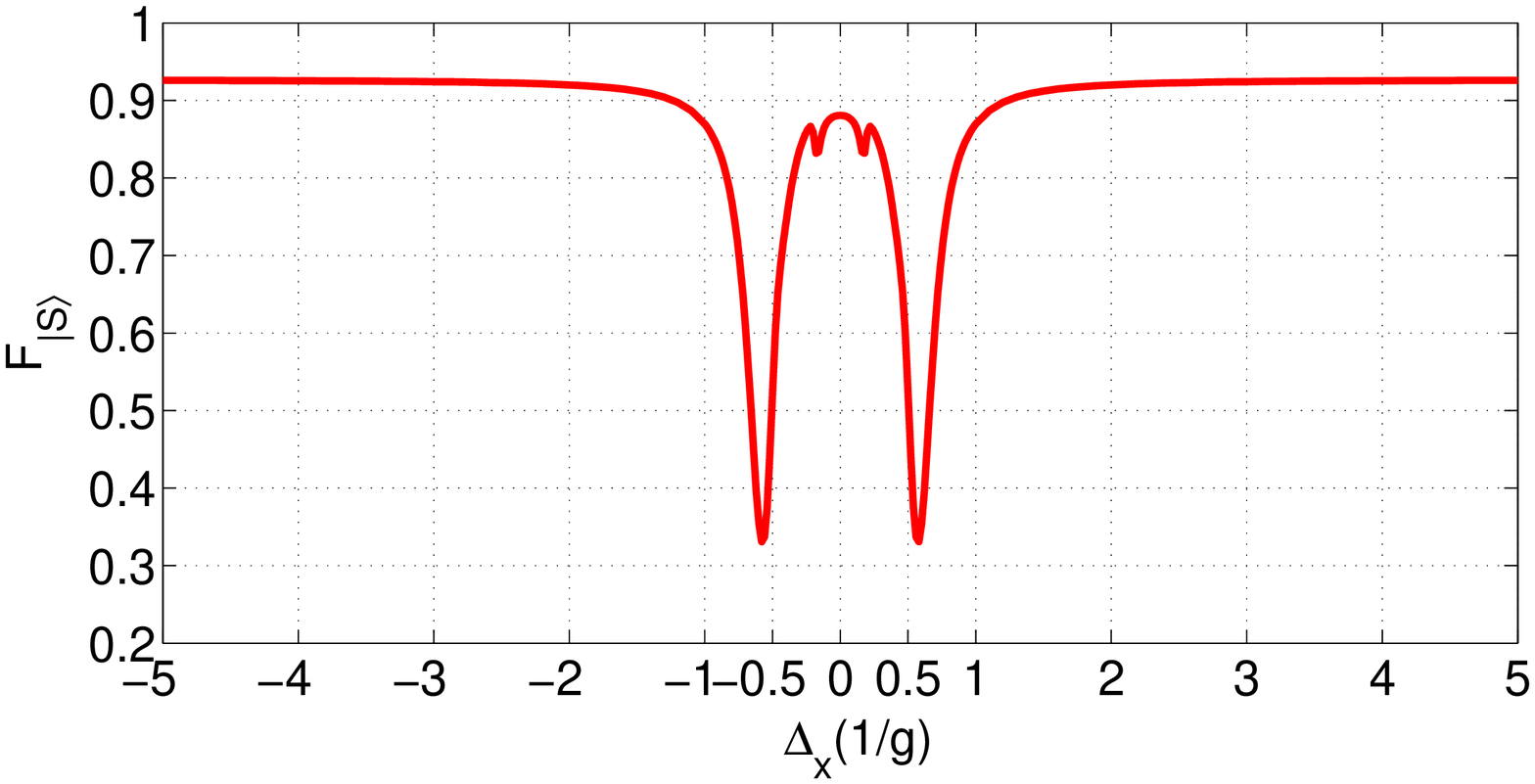}}
\caption{(Color online) Fidelity versus frequency spacing of the
mediating bosonic modes. The parameters are the same as those in
Fig. 3: (a) $N=2$; (b) $N=3$; (c) $N=5$.}
\end{figure}
\begin{figure}\label{fig-7}
\centering
\includegraphics[width=0.5\columnwidth]{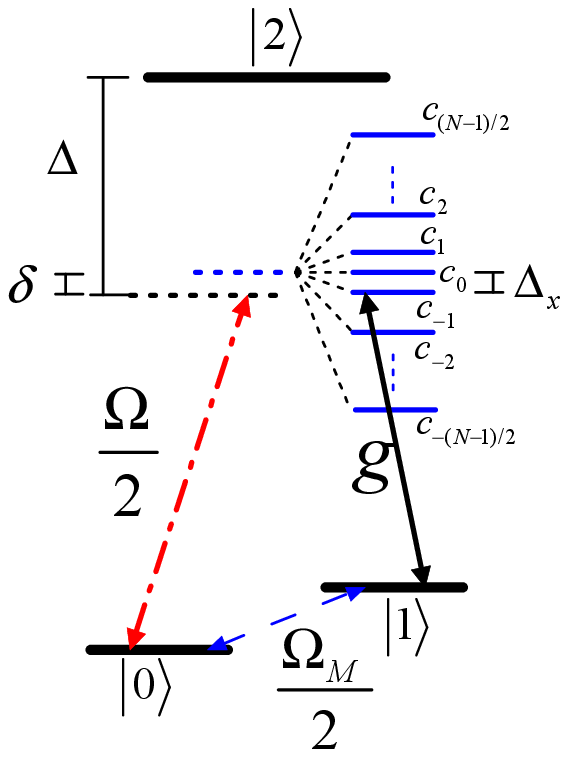} \caption{(Color
online) The effective atom-field coupling when $N$ mediating modes
are involved. The couplings between two cavity modes and $N$
mediating modes lead to $N+2$ delocalized field modes (blue solid
line) with different frequency spacings, which interact with two
atoms respectively. $c_{0}$ is the delocalized field mode resonant
with the cavity modes. $c_{1}$ and $c_{-1}$ are the neighboring
modes nearest to $c_{0}$, while $c_{-(N-1)/2}$, ..., $c_{-2}$ and
$c_{2}$, ..., $c_{(N-1)/2}$ are other delocalized field modes.}
\end{figure}
Numerical simulations show that the fidelity of the steady-state
entanglement is free from the effects caused by dispersive mediating
modes when $\Delta_{x}$ $>$ $\delta$. Strong coupling for photons
between distant nanocavities mediated by a waveguide at room
temperature has been reported and the system dynamic control can be
realized \cite{NPHOTONICS2011-6-56}. The cooperativity parameter $C$
about $100$ has been realized \cite{Nature2007-445-896}. Based on
these experimental techniques, the present scheme is feasible.

\section{Conclusions}
In summary, we have proposed a dissipation based scheme by which two
atoms trapped in two distant cavities can be driven to steady-state
entanglement. The competition based on the unitary dynamics and the
dissipative dynamics leads to the entangled steady-state. Our
results show that the dissipative mediating bosonic mode can be used
as an entanglement catalyst for two distant atoms. The effects of
dispersive mediating modes are analyzed. It is shown that the
entanglement is robust against the fluctuations of the Rabi
frequencies of the classical fields. We show that the distributed
steady-state entanglements can be obtained with high fidelity
regardless of the initial state and there is a linear relation in
the scaling of the fidelity with the cooperativity parameter.

L.T.S., X.Y.C., H.Z.W. and S.B.Z. acknowledge support from the Major
State Basic Research Development Program of China under Grant No.
2012CB921601, National Natural Science Foundation of China under
Grant No. 10974028, the Doctoral Foundation of the Ministry of
Education of China under Grant No. 20093514110009, and the Natural
Science Foundation of Fujian Province under Grant No. 2009J06002.
Z.B.Y acknowledges support from the National Basic Research Program
of China under Grants No. 2011CB921200 and No. 2011CBA00200, and the
China Postdoctoral Science Foundation under Grant No. 20110490828.

\end{document}